\documentclass[letters, fleqn,usenatbib]{mnras}

\usepackage{verbdef}
\usepackage{siunitx}
\usepackage{gensymb}
\usepackage{comment}
\usepackage[normalem]{ulem}
\usepackage[svgnames]{xcolor}
\verbdef{\vtext}{verb text}
\usepackage{xcolor}

\usepackage[T1]{fontenc}

\DeclareRobustCommand{\VAN}[3]{#2}
\let\VANthebibliography\thebibliography
\def\thebibliography{\DeclareRobustCommand{\VAN}[3]{##3}\VANthebibliography}

\usepackage{graphicx}	
\usepackage{amsmath}	
\usepackage{amssymb}	
\usepackage{newtxtext,newtxmath}

\title[The Impact of Cluster Mergers in the Red Sequence.]{Clash of Titans: The Impact of Cluster Mergers in the Galaxy Cluster Red Sequence.}

\author[Franklin Aldás]{Franklin Aldás,$^{1}$\thanks{E-mail: franklin.aldas@userena.cl}
Alfredo Zenteno,$^{2}$
Facundo G\'omez,$^{1,3}$
Daniel Hernandez-Lang,$^{4,5}$\newauthor
Eleazar R. Carrasco,$^{6}$
Cristian A. Vega-Martínez,$^{1,3}$ \&
J. L. Nilo Castellón,$^{1}$
\\
$^{1}$Departamento de Astronomía, Universidad de La Serena, Avenida Juan Cisternas 1200, La Serena, Chile\\
$^{2}$Cerro Tololo Inter-American Observatory, NSF's NOIRLab, Casilla 603, La Serena 1700000, Chile\\
$^{3}$Instituto Multidisciplinario de Investigación y Postgrado, Universidad de La Serena, Raúl Bitrán 1305, La Serena, Chile \\
$^{4}$Faculty of Physics, Ludwig-Maximilians-Universit\"{a}t, Scheinerstr.\ 1, 81679 Munich, Germany \\
$^{5}$Excellence Cluster Origins, Boltzmannstr.\ 2, 85748 Garching, Germany\\
$^{6}$Gemini Observatory, NSF's NOIRLab, Casilla 603, La Serena 1700000, Chile.
}

\date{Accepted XXX. Received YYY; in original form ZZZ}

\pubyear{2023}

\begin{document}
\label{firstpage}
\pagerange{\pageref{firstpage}--\pageref{lastpage}}
\maketitle

\begin{abstract}
Merging of galaxy clusters are some of the most energetic events in the Universe, and they provide a unique environment to study galaxy evolution. We use a sample of 84 merging and relaxed  SPT galaxy clusters candidates, observed with the Dark Energy Camera in the $0.11<z<0.88$ redshift range, to build colour-magnitude diagrams to characterize the impact of cluster mergers on the galaxy population.
%
We divided the sample between relaxed and disturbed, and in two redshifts bin at $z = 0.55$.
%
%
When comparing the high-z to low-z clusters we find the  high-z sample is  richer in blue galaxies, independently of the cluster dynamical state.
In the high-z bin we find that disturbed clusters exhibit a larger scatter in the Red Sequence, with wider distribution and an excess of bluer galaxies compared to relaxed clusters, while in the low-z bin we find a complete agreement between the relaxed and disturbed clusters.
%
%
%
%
Our results support the scenario in which massive cluster halos at $z<0.55$ galaxies are quenched as satellites of another structure, i.e. outside the cluster, while at $z \geq 0.55$ the quenching is dominated by in-situ processes.


\end{abstract}




\section{Introduction}

Galaxy clusters are the largest collapsed gravitationally bound structures of the Universe and typically contain  hundreds or thousands of 
galaxies~\citep{Voit2005, Abell1954}. These clusters have grown to their present-day mass through the merger and accretion of neighbouring substructures. Galaxy clusters can be found in different dynamical states. In  a relaxed cluster, the Intra-Cluster Medium (ICM) is generally close to thermodynamic equilibrium. However, in clusters colliding with other massive structures, the released binding energy of the systems results in heating of their ICM \citep{Sarazin2002}. Cluster mergers are the most energetic events after the Big Bang and  can emit energies up to  $10^{64}$ erg \citep[][]{sarazin04, Kravtsov2012}. Indeed, observations in different wavelengths suggest that a significant fraction of clusters, between 30\% and 70\%,   are not fully virialized \citep{Dressler1988, Hou2012, Zhang2009,  Andrade-Santos2017}.   Such energetic cluster merging events can trigger and/or quench star formation in their galaxy members, thus affecting their observable properties such as colours.

 The dynamical state of galaxy clusters is described using different observational proxies. Among them we find the central cooling time~\citep{Bauer2005}, the concentration index which is the excess of the  surface brightness in the central part of the cluster measured in X-rays~\citep{Santos2008, Yuan2022}, the displacement between the the cluster X-ray surface brightness peak from its surface brightness centroid~\citep{Cassano2010, Yuan2022}, and, combinations of different observations such as the offset between the position of the brightest cluster galaxy (BCG) and the Sunyaev-Zeldovich (SZ) centroid~\citep{ Zenteno2020a}. Additionally, some quantities can be used  to study the dynamical state using cosmological simulations, for example, the virial ratio $\eta$ based on the virial theorem that measures the degree of deviation that a cluster has from an equilibrium state, the fraction of mass of the cluster contained in subhalos with respect to its total mass, and the offset between the centre of mass of cluster with the cluster centre, or a weighted combination of those parameters~\citep{Zang2022}.

Today, a significant fraction of the members inhabiting these clusters consist of elliptical and lenticular galaxies \citep{Hubble1931, Oemler1974,  Dressler1980}. 
These early-type galaxies feature a well defined linear relation between their colour and magnitude \citep{Visvanathan1977, Bower1992, Kodama1997}, the so called red sequence (RS). Such galaxies in the cluster RS  have null or little ongoing star formation \citep[][]{Gladders2004,Gladders2007,DePropris2016}, and their colour evolution can be remarkably well described by simple evolutionary models \citep{Stanford1998}. In fact, such models are so successful that they have been used to identify clusters \citep{Gladders2000,Murphy2011,Bleem2015,rykoff16, Vakili2020} and to provide robust photometric redshifts up to $\sim$ 1.5  \citep[e.g.][]{Song2012, Bleem2015, bleem20}, with a precision better than $\sim 0.01 \times (1 + z)$ up to z $\sim$ 1.0 \citep[e.g.][]{rykoff16, klein18, klein19}. 
%
%
While the slope and zero-point evolution of the RS  describes the aging of galaxies’ stars formed at $z\sim$2 - 3 \citep[at least up to z $\sim$1.3, e.g.,][]{mancone10}, the  scatter provides information about its stellar population age diversity~\citep{Connor_2019}. 

Previous studies have shown that the scatter remains relatively constant up to z $\sim$1  \citep[e.g.][]{jaffe11, Hennig2017}. However, at $z\gtrsim 1.3$,  the RS is found to be wider and bluer, indicating that at such $z$ clusters are approaching their star formation epoch \citep{hilton09,papovich10,snyder12}.  Furthermore, \citet{brodwin13} suggest that $z \sim 1.4$ could be where clusters star formation activity ends and the era of passive evolution begins.

Previous works have studied the differences in the star formation rates of galaxies in disturbed clusters concerning their relaxed counterpart. Yet, there is no agreement about whether a galaxy's star formation rate is stimulated or suppressed  during large merger events.
%
For example, \citet{Pranger2014} through an analysis of the galaxy population
in Abell 3921, and \cite{Kleiner2014} in A1750, found that  mergers quench star formation. \citet{Shim2011} also studied the interacting cluster A2255 finding that the merging process suppresses star formation and transforms galaxies into quiescent galaxies.
However, other authors such as \citet{Ferrari2003}, by an spectroscopic analysis of Abell 521, and \cite{Owers2012} by studying the cluster Abell 2744 (which is currently undergoing a major merger) found that the high-pressure merger environment triggers star formation. Supporting this,  ~\cite{Sobral2015} and ~\cite{Yoon2020} found that the star formation rate in interacting clusters is around ~20\% higher than the observed in  relaxed structures.  Such findings are confirmed using H$\alpha$ observations of disturbed and relaxed clusters at $z<0.4$ 
finding a higher prevalence of H$\alpha$ emitter galaxies in disturbed clusters, within 2 Mpc from the cluster's center, than in relaxed clusters \citep{Stroe2017,stroe21}. In the same way, \citet{Hou2012} studying groups of galaxies 
at intermediate redshifts ($z\sim 0.4$)  show that galaxies in groups with substructures present a significantly higher blue galaxies population 
compared to galaxy groups with no detected substructures.

The goal of this work is to study the impact that the merging of clusters has on their red galaxy population. 
%
We compare the cluster galaxies RS of a sample of 84 clusters, both relaxed and disturbed, all within a wide redshift range  between 0.11$\leq z \leq $0.88.  The cluster sample was first introduced in \citet{Zenteno2020a}. 
The paper is organized as follow: in \S~2 we provide details of the observations and data reduction. In \S~3 we show the data calibration between DES and Munich pipeline reductions, while in \S~4 we report our findings. Finally in \S~5 are our conclusion. Throughout the paper we assume a flat Universe, with a $\Lambda$CDM cosmology, $h$ = 0.7, $\Omega_M$= 0.27 \citep{komatsu11}. 

\section{Data}
The observations used in this paper  were carried out with the Dark Energy Camera \citep[DECam;][]{Flaugher2015}, a 570 Mega-pixels CCD, installed
at the prime focus of the V. M. Blanco 4-meter Telescope at Cerro Tololo Inter-American Observatory (CTIO). The DECam 
has a  field-of-view  of $2.2\degree$. 
We use two sets of data; 
65 clusters ($0.11 \lesssim z\lesssim0.65$) comes from  the Dark Energy Survey public second data release  \citep[DES; ][]{DES2021}, while for  19 clusters at $z\gtrsim0.65$ we use archival data as well as data obtained using Director Discretionary Time (DDT). DDT was pooled with the DECam eROSITA Survey (DeROSITAS;  PI Zenteno) allocation,  taking advantage of the flexibility a large pool of nights provide to programs with different needs.  

%


DES is a 5000 square degree optical survey using the DECam and 5 filters \verb|g,r,i,z,Y|, covering a wavelength range from $400$ nm to $1065$ nm.  The Data Release 2 (DR2) of DES is the result of six years of observations (2013-2019) collecting information of around 700 million galactic and extragalactic sources \citep{DES2021}.  The image reduction  and processing for the DES sample set were done by the DESDM system. This process performs flat-fielding, bad-pixel masking, overscan removal, masking of cosmic rays and artificial satellites, and other image corrections \citep{Morganson2018}. Once the images are fully reduced, the pipeline performs a fitting with PSFEx and source detection using Source Extractor generating the 
co-added images and its associated source catalogues that are ready for science analysis \citep{DES2021, Bertin2011, Bertin1996}.  
We retrieved the source catalogues from the DES DR2 repository. 
The  downloaded 
parameters are: \verb|MAG_AUTO, MAGERR_AUTO, FLAGS, IMAFLAGS_ISO|, and \verb|SPREAD_MODEL| for each  $g$, $r$, $i$, $z$ bands. 
\verb|MAG_AUTO| are the magnitude estimations using an elliptical model considering the Kron radius, with \verb|MAGERR_AUTO| their uncertainties~\citep{Kron1980}. \verb|FLAGS| store additive flags indicating potential problems in the source extraction process, and \verb|IMAFLAGS_ISO| are flags where the sources have missing/flagged pixels in their single epoch images. Finally, the \verb|SPREAD_MODEL| is a parameter to identify extended sources comparing the fit quality between the local point-spread function (PSF) and an extended circular exponential disk \citep{desai12, DES2021}. 
The limiting magnitudes of the DES (DR2) for the selected clusters, at $10 \sigma$ are $g\sim 23.7$, $r\sim23.7$, $i\sim23.0$, and $z\sim22.27$ mag.

DeROSITAS is a 
survey designed to complement the German sky of the eROSITA survey \citep[][]{Merloni12} in the optical wave range. DeROSITAS was performed using DECam in filters $g$, $r$, $i$, and $z$, reaching minimum
depth of 22.7 (23.5), 23.2 (24.0), 23.3 (24.0), 22.5 (23.2) AB magnitudes at 10(5)$\sigma$.  DeROSITAS observing strategy consisted in filling the sky avoiding archival data when at sufficient depth, and carry out observations in coordination with other current surveys, such as DELVE \citep[][]{Drlica-wagner21}, to avoid duplication.  
During DeROSITAS nights, high-z clusters  observations were triggered when the seeing and the effective time t$_{\rm eff}$ \citep[][]{bernstein16} were better than average (seeing better than $\sim$1.0'' and t$_{\rm eff} > 0.4$). 
The t$_{{\rm eff}}$ is a scale factor to be applied to the open shutter time to reflect the quality of the observations compared to good canonical conditions. These good  conditions are defined as observations with a FWHM of 0.9" and sky brightness obtained when pointing the telescope to the zenith under dark conditions.     

The DeROSITAS observations used here reach magnitudes $i\sim 23.9$, and $z\sim 23.6$, which is between 0.9 and 1.3 magnitudes deeper than DES (we used just those bands because those observations are focused in high-redshift clusters). The data reduction was done using a pipeline similar to DESDM \citep{desai12}, where the steps are done by first building single epoch (SE) images and then using a co-adding pipeline. The single epoch pipeline groups the observations according to the observation night and then, for each DECam observation that contributes to the cluster area (within $1\times1$ deg$^2$ from the SPT position), it constructs $\sim$62 (one for each CCD) photometrically flattened, astrometrically calibrated single SE images, together with position variable PSF models and PSF corrected model fitting catalogs. The processing includes overscan and bias correction, flat-fielding, initial astrometric calibration and PSF corrected model fitting photometry using PSFex. Final astrometric and photometric calibrations for each SE image are done using Gaia DR2 \citep{Evans2018} photometry data \citep[for details refer to][]{George2020a}. Finally, the coadd pipeline works similar to the DES processing, generating PSF homogenised COADD images and catalogues by using a combination of SourceExtractor and PSFex softwares.

The final {\it high-z clusters} sample is presented in the Table~\ref{DeROSITA}, and the DES cluster sample is presented in the Table~\ref{DES}. 

\section{Catalogs}
\label{sec:catalogs}
\begin{figure}
\centering
\includegraphics[width=\linewidth, trim={1cm 0.5cm 2.5cm 1cm}, clip]{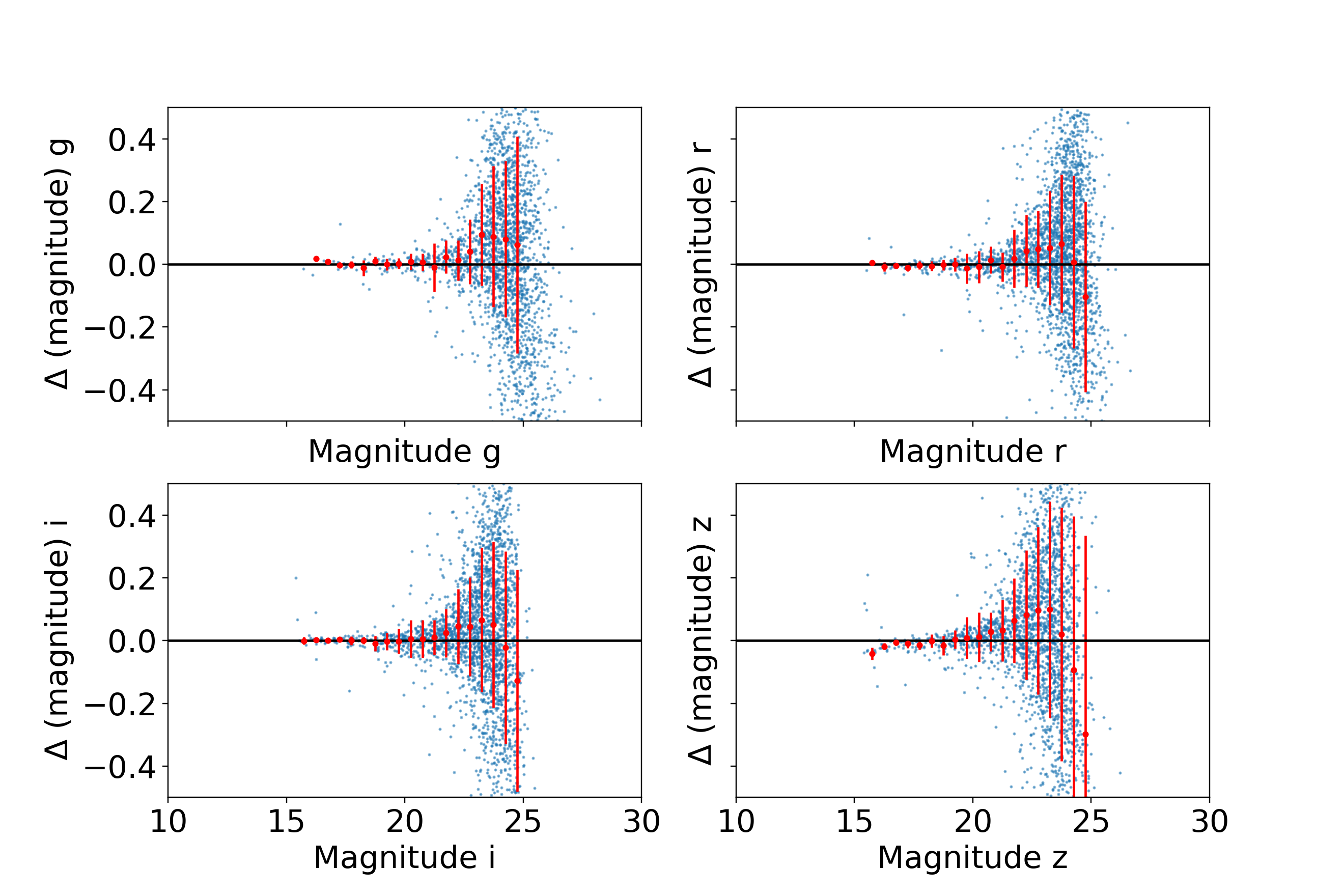}
\caption{Example of the difference between magnitudes measured by DES and {\it high-z clusters}, in each used filter for stars detected in the field of the cluster SPT-CLJ0310-4647, located at $z=0.710$ after the SLR correction. The red dots is the average difference between those MAG$\_$AUTO and the error bars correspond to the standard deviation of the differences.}
\label{fig:calibration_final}
\end{figure}
As we mentioned, we have two sets of data. The first comes from the DES DR2 database public repository, and the second, with deeper photometry, comes from our own reductions. Following the catalogues calibration described in the preceding section, hereafter we will create the final sample joining both catalogues in the following way: 65 clusters from DES described in Table \ref{DES} at $0.1<z<0.65$ (henceforth the {\it DES clusters} sample), and the 19 clusters from DeROSITAS at redshift higher than 0.65 (henceforth the {\it high-z clusters} sample) detailed in Table \ref{DeROSITA}.   
%
As those two catalogues were obtained from two different pipelines, we expect slight differences.  To reduce such photometric differences we correct the colour of both catalogs using the Stellar-Locus-Regression (SLR) technique \cite{High2009}.  Once the colour is corrected we adjust the zeropoint by adopting the DES zeropoint for the {\it high-z clusters} catalogs. The process is outlined below:

First, we calibrated the colours $g-r$, $g-i$, $r-i$, and $i-z$ for both sets of data using the SLR code. This technique uses a region in the colour-colour diagram populated by stars~\citep{Covey2007, Ivezic2007}. The SLR code accurately calibrates  the colours for stars and galaxies using catalogued flat-fielded images  without having to measure standard stars or  determining the zero-points for each pass-band.  
The SLR technique also corrects for differences in instrumental response,  atmospheric response, and for  galactic extinction. 

As input, the SLR code needs magnitudes, magnitudes errors, and extinction value for every source in each passband used ($g$, $r$, $i$, $z$). 
The dust extinction were obtained using the Schlafly \& Finkbeiner Dust \citep[]{Schlafly2011, Schlegel1998}.  
Colour correction was made considering objects classified as stars. 
The photometric catalogues include the \verb |SPREAD_MODEL| parameter, which is a star-galaxy separator.  
Following the same criteria used by ~\citet{Hennig2017}, we consider as stars the sources with  \verb |SPREAD_MODEL < 0.002|. We clean the final sample by excluding sources with \verb|IMAFLAGS_ISO > 0| in all bands to avoid saturated objects and objects with missing data ~\citep{Morganson2018} and \verb|FLAGS| $\geq$ \verb|4|, to include deblended sources but excluding sources flagged with warnings during the extraction process ~\citep{DES2021}. 


As a result of the  SLR  calibration, we obtained 
corrected colours for stars and galaxies for the DES and {\it high-z clusters} catalogs.  Next, we correct the absolute magnitude of the {\it high-z clusters} catalogs by comparing its star's magnitudes to the DES star magnitudes for the $i-$band.  We used a 0.25 arcseconds to match stars and a 0.5 magnitude difference to avoid variable stars.  Once such correction is found (which has an average value of $\sim$ 0.02 magnitude), we applied the correction to all objects in the {\it high-z clusters} catalog.  

As an example of the results of the calibration process,
in Figure \ref{fig:calibration_final} we present four diagrams, one for each band ($g$, $r$, $i$, $z$)  for the catalogues corresponding to the cluster SPTCLJ0310-4647. In those diagrams, each point corresponds to a Milky-Way star detected in the field of view of the cluster,  present in both DES and DeROSITAS catalogues. Stars were selected by  \verb|SPREAD_MODEL < 0.02|). 
In the x-axis, we show the magnitude (MAG$\_$ AUTO) corresponding to the DES catalogue (taken as a base to calibrate the high-redshift data set),meanwhile in the y-axis we have the $\Delta (magnitude) $ defined as the difference between DES and {\it high-z clusters} MAG$\_$AUTO.  Similar plots were obtained for the 19 clusters in common. 
We can see that the differences in magnitude for DES and {\it high-z clusters} observations  are close to zero for the brightest end. The maximum mean difference for the brightest stars (mag < 20.5), between the two catalogues, computed as the average difference between the DES and DeROSITAS magnitudes, is around 0.04 mag in all four bands.  This result suggests that the fluxes obtained using both pipelines are very close  to each other and, thus, we can consider them as equivalents.
\begin{figure}
\centering
\includegraphics[width=0.95\columnwidth, trim={1cm 0.5cm 3cm 2.5cm}, clip]{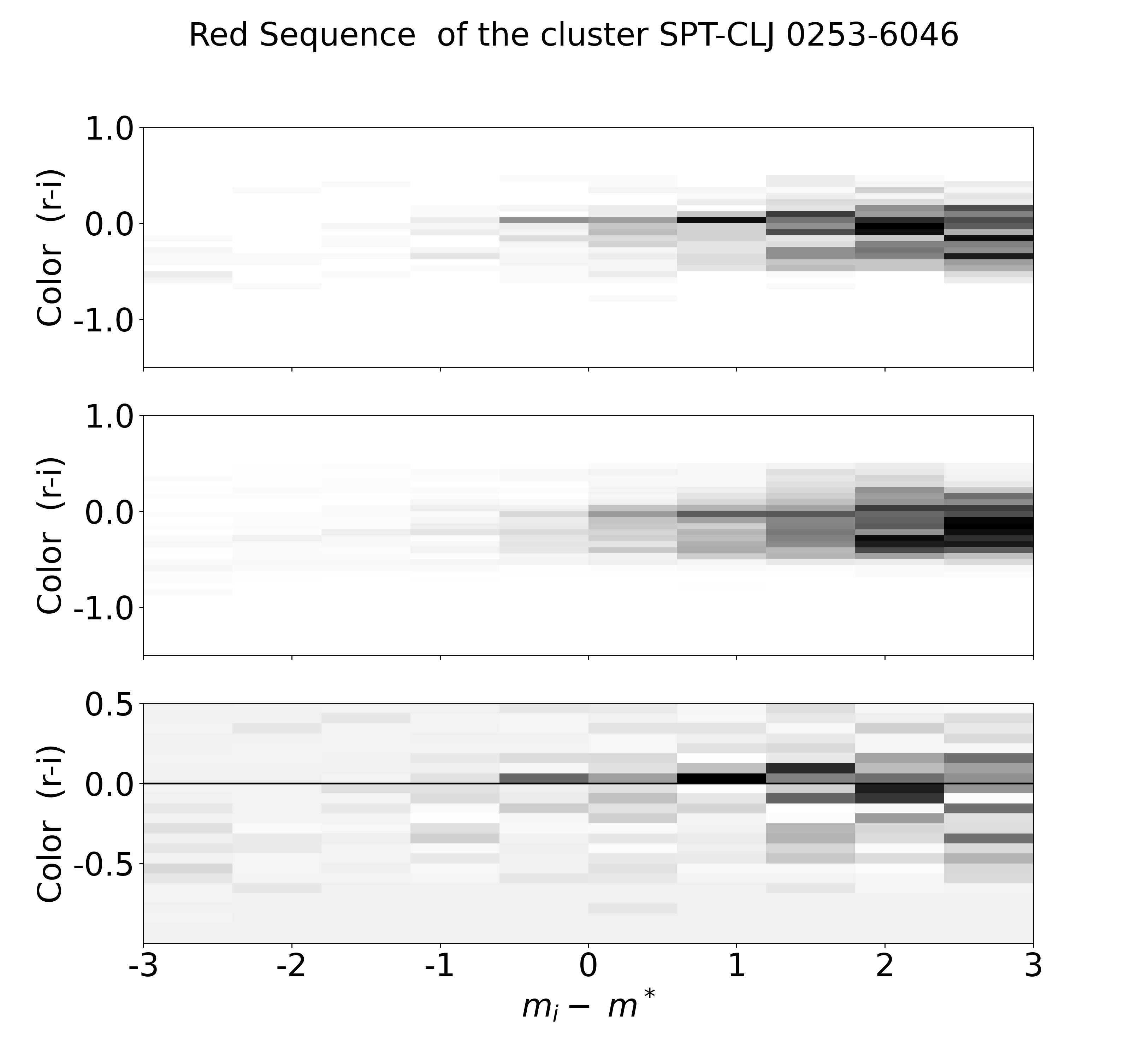}
\caption{Background correction process for the cluster SPT-CLJ0253-6046. On the top panel, we have the detected objects inside the virial radius $R_{200}$. In the middle panel, the objects found in the background, located in the annulus between $1.5-3 R_{200}$. Finally, in the bottom panel, the Red Sequence of the cluster computed as the overdensity corrected by the background. The three panels were plotted after performing the redshift slope correction for the RS time evolution. The used bins for the CMDs have a size of 0.6 in magnitude and 0.06 in colour. }
\label{fig:Red_sequence_cluster}
\end{figure}
\section{RESULTS}
The goal of this paper is to study the differences in the cluster RS galaxy population as a function of the cluster dynamical state. 
A galaxy spectrum is mostly flat and is mainly composed  of a combination of blackbody emitters, but there is a noticeable break at 4,000 \AA\ in the rest frame where there is an absorption of high energy radiation in the stellar atmospheres in metal poor populations \citep{Mihalas1966, Mihalas1967, Poggianti1997}. This break allows to  separate  blue from red galaxies~\citep{Poggianti1997}. For this reason, we used two photometric bands that contain the  Balmer break in the rest frame. In those filters, the elliptical galaxies tend at a given redshift range, tend to be  redder than normal  galaxies at any lower redshift, thus becoming easily noticeable from the background \citep{Gladders2000}. 
Then, the selected bands depend on the redshift of each cluster, and are presented in the Table~\ref{table:bands_redshift}.

\begin{table}
\centering
\caption{Bands used for the colour- magnitude diagram depending on the cluster redshift to capture the 4000 {\AA} Balmer break. }
\label{table:bands_redshift}
\begin{tabular}{ccc}
\hline
Redshift & color  bands & magnitude band \\
\hline
$0<z\leq0.33$ & (g-r) & r\\
$0.33< z \leq 0.74$ & (r-i) &i \\
$0.74< z \leq 0.9$ & (i-z) &z\\
\hline
\end{tabular}
\end{table}

\begin{figure*}
\centering
\includegraphics[width=\linewidth]{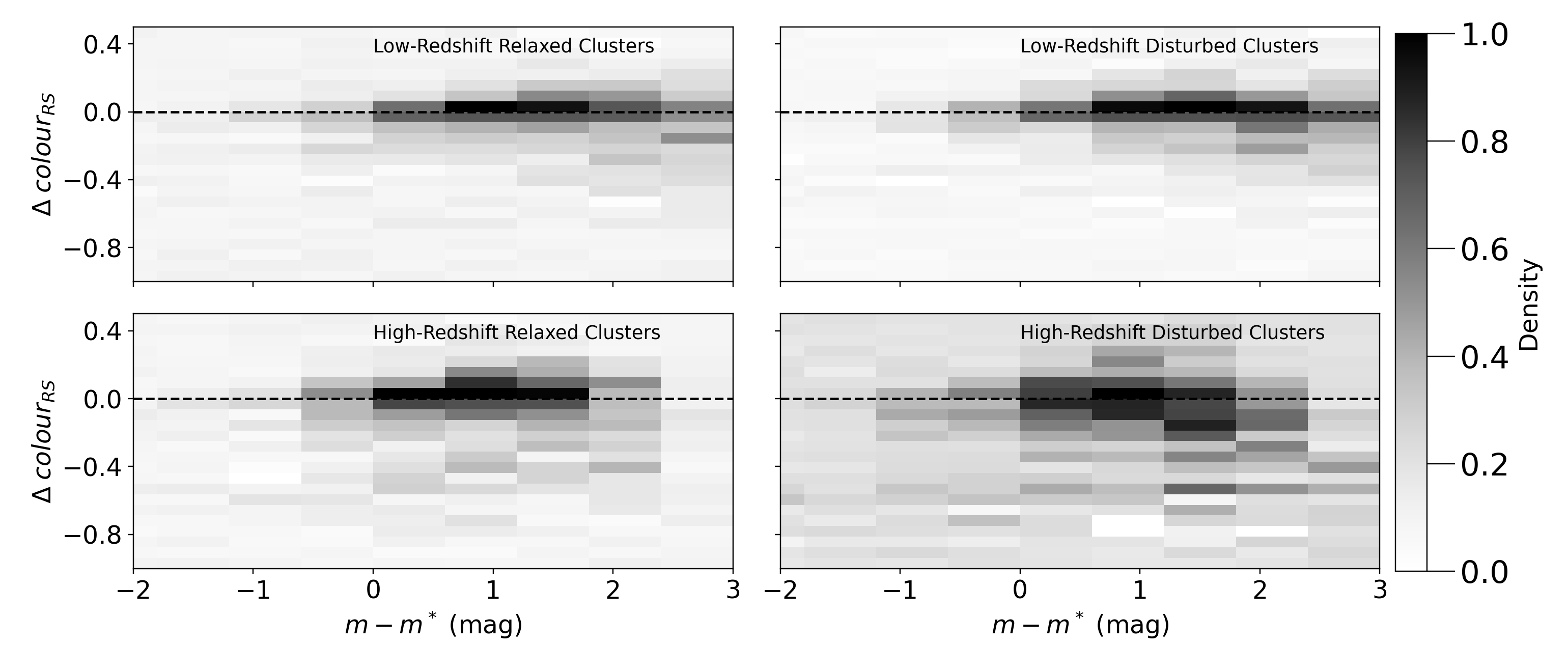}
\caption{Colour magnitude diagram of galaxies for each studied clusters subsets. In the top row we have the low -redshift $(z<0.55)$ relaxed (left) and disturbed clusters (right) and in the bottom row we have the relaxed (left) and disturbed (right) clusters for the high redshift bins $(z>0.55)$.}
\label{Fig:Red_Sequence}
\end{figure*} 
\begin{table}
\centering
\caption{Number of clusters in the used sample in each analyzed group. We divided the sample between low and high redshift samples and relaxed and disturbed clusters.}
\label{table:number_clusters}
\begin{tabular}{ccc}
\hline
Redshift range & Relaxed clusters & Disturbed clusters \\
\hline
$0.1<z<0.55$ & 16 & 27 \\
$0.55\leq z<0.9$ &  25 & 16\\
\hline
\end{tabular}
\end{table}
To construct a cluster colour-magnitude diagram (CMD), we consider  galaxies within $R_{200}$  as cluster galaxies and galaxies in the annulus between $1.5\times R_{200}< r < 3\times R_{200}$ as background. Where $R_{200}$ is defined as the radius where the cluster density is 200 times the critical density of the Universe at a given redshift. The cluster centers corresponds to the SPT-SZ centers \citep{Bleem2015}, and the $R_{200}$ were estimated by \citet{Zenteno2020a} using the estimated $M_{500}$, and the Duffy mass-concentration relation \citep{Duffy2008}. We bin both the cluster and background CMDs, using bins of 0.6 in magnitude and 0.06 in colour, and perform a statistical background subtraction to correct for contamination due to the projection effects. 

Once the background-subtracted CMD is built for each cluster, we use a stellar population synthesis models to stack them. We do this by using models with an exponential starburst decay and a Chabrier IMF  \citep{Bruzual2003} for six metallicities described in \citet[e.g.,][]{Song2012}, and \citet{zenteno16}. 
Using the cluster's redshift, and following the procedure 
as described by \citet{Hennig2017},  we obtain the model (expected) cluster RS slope as a function of the magnitude and then subtract it from the observed slope, bringing the cluster's colour slope to zero. The model also provides the characteristic magnitude defining the 
knee of   the luminosity function, $m^*$, for each filter and redshift. The cluster redshift used in this paper corresponds to the same as \citet{Zenteno2020a}, which are photometric and spectroscopic redshifts collected from several literature sources.  \\
In Figure \ref{fig:Red_sequence_cluster}, we show the process of the background correction for the SPT-CLJ0253-6046 galaxy cluster, located at $z= 0.45$. For this cluster we used the $r$ and $i$ bands. In the horizontal axis we have the magnitude ($m_i-m^*$), with $m^* $ obtained from the model. The three vertical axes show the $r-i$ colour.     In the top panel, we show the density of objects within the virial radius, plotted in gray scale. The middle panel focuses on objects located between $1.5 $ to $3$ viral radius. Finally, in the bottom panel we show the resulting red sequence, obtained after subtracting the number object in the background from the number of objects within $R_{200}$,  normalized by their areas.

%
%
We then combined the cluster CMDs by adding all background-corrected numbers of galaxies, normalised by the total number of clusters used in each bin.
\citet{DePropris2016} studied the evolution of galaxies as they experience gravitational infall into cluster cores during merging processes, resulting in a transformation of the cluster RS (red sequence) morphology, as previously observed by~\citet{DeLucia2007, Stott2007}. The RS evolution below $z \sim 0.5 - 0.6$  shows that there are little or no morphological changes in the galaxy population, unlike earlier in time. For this reason, we separated the cluster sample in two redshift bins: the first one includes clusters between $0.1<z<0.55$ (low redshift bin), and the second one includes clusters with redshifts between  $0.55\leq z < 0.9$. Additionally, we subdivided our cluster sample according to their dynamic state. As a result, we have four subgroups of clusters: disturbed and relaxed clusters at low and high redshifts. The number of clusters in each subgroup is presented in Table~\ref{table:number_clusters}. It should be noticed that we have at least 16 clusters in each subgroup. 

The dynamical state of the clusters used in this paper was estimated by~\citep{Zenteno2020a}. They classified between relaxed and disturbed clusters using four different proxies: The offset between the position of the BCG and the SZ centroid ($D_{BCG-SZ}$), the core temperature, the morphological parameter($A_{Phot}$), which measures the asymmetry of the X-ray emission, and the offset between the BCG and the peak of X-ray emission. In this analysis, a cluster is defined as relaxed if it met any of the following three conditions: i) the cluster has $A_{phot} < 0.1$, ii) the cluster has a cool core $(K_0<30~\text{keV cm}^{2})$~\citep{McDonald2013}, or iii) the offset between BCG and X-ray peak is less than 42 kpc~\citep{Mann2012}. Meanwhile, the clusters were classified as disturbed if the offset between the position of the BCG and the position of the SZ centroid $D_{BCG-SZ}$ is greater than 0.4 $R_{200}$. The last criterion was chosen given that the distribution of  $D_{BCG-SZ}$  looks flat after this value. Clusters that don’t meet any of the four previously mentioned criteria are considered in an intermediate evolutionary state and are excluded from this analysis. The BCG position is used as a proxy of the collisionless component since it is expected to quickly fall to the lowest region of the potential well~\citep{Tremaine1990}. This position was derived from optical observations. Meanwhile, the centroid of the SZ was used as a proxy for the collisional component. When a cluster interacts with other clusters or groups, the gas, dark matter and galaxy components act differently depending on their nature. Then, the offset between the collisional and collisionless matter components can be used as a proxy to quantify the relaxation state of clusters~\citep{Zenteno2020a}.


In Figure~\ref{Fig:Red_Sequence}, we present the stacked CMD for the four subgroups: low and high redshift 
relaxed clusters and low and high redshift 
disturbed clusters. The gray scale covers the same numeric range in the four panels. Darker (lighter) colours represents higher (lower) density regions in  colour-magnitude space. In all panels we see a dominant RS with an associated bluer galaxy population.  For the low redshift sample, we find that disturbed and relaxed clusters red sequence show  similar colour distributions. 
This is, the relative contribution from  early- and late-type galaxy populations to the CMD is comparable in both cases, indicating that the current dynamical state of a galaxy cluster has little impact on its CMD at the present-day. 
%
On the other hand, on the bottom panels of Figure \ref{Fig:Red_Sequence} we show the RS for the high redshift sample. 
These RS distributions are generally wider with respect to the low-redshift counterpart. 
More interestingly, we observe a wider colour distribution in the disturbed cluster population when compared with both the high redshift relaxed distribution and the overall low redshift sample.

%
%

\begin{figure}
\centering
\includegraphics[trim={0.0cm 1cm 1cm 2.5cm}, clip, width=\linewidth]{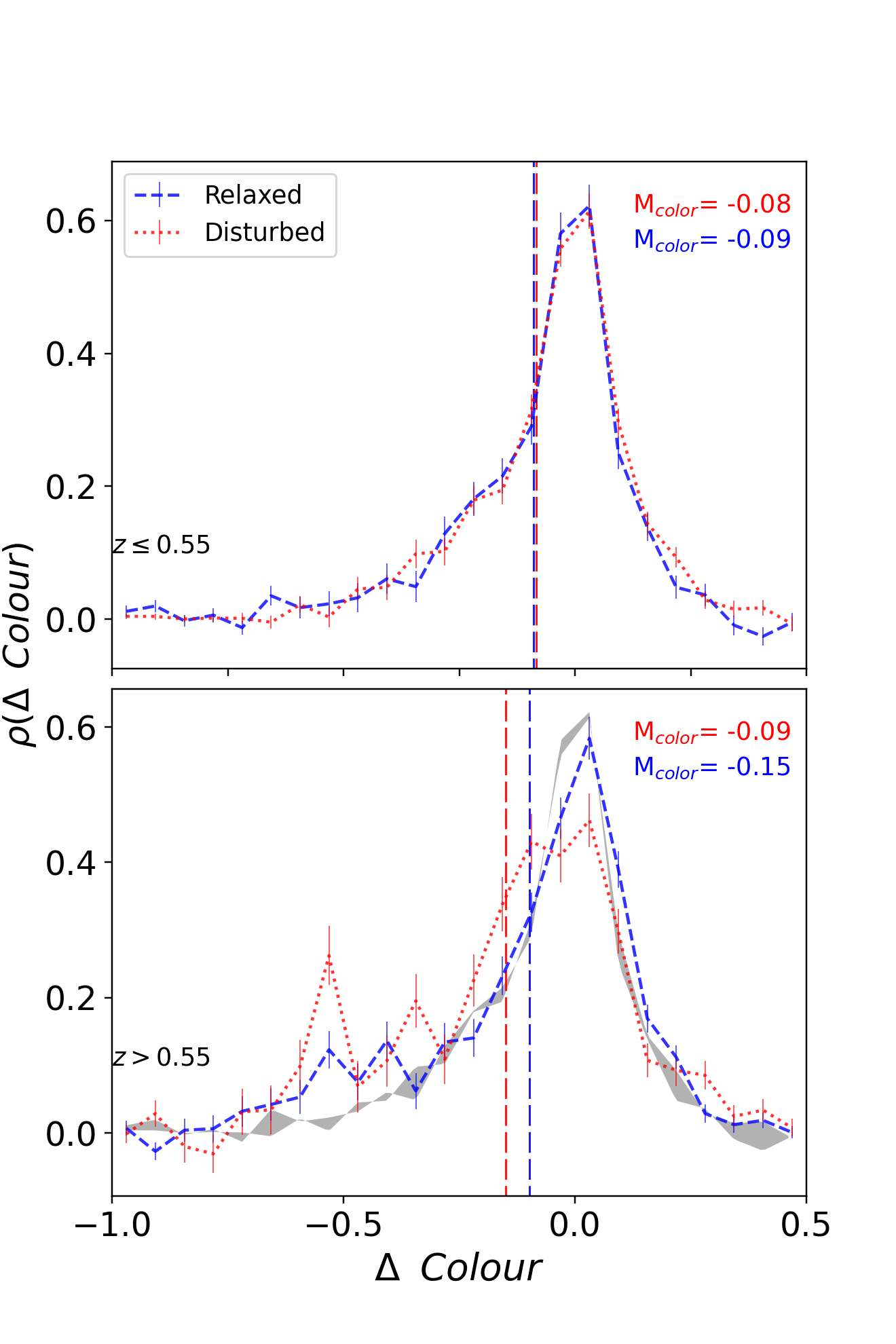}
\caption{Colour density distributions for galaxies in each sample bin. The top panel presents the colour density distribution for the low-redshift bins relaxed in the blue line and disturbed in the red line and their error bars were computed as a Poisson noise. \textcolor{blue}{The  vertical dashed lines indicate the median of the corresponding distribution}.  The bottom panel presents the same as described before,  but for the high redshift $(z>0.55)$ clusters, also in gray scale we show the density distribution for the low-redshift clusters.}
\label{Fig:Density}
\end{figure}

Figure~\ref{Fig:Density} shows histograms from collapsing Figure~\ref{Fig:Red_Sequence}  in the magnitude ($m-m^*$) axis for the four studied subgroups. Those histograms are normalized by their euclidean norm.  In the upper panel we have the relaxed and disturbed clusters for low redshift clusters ($z<0.55$). Similarly, in the bottom panel we have the galaxy color distribution for high redshift clusters ($z>0.55$),  relaxed in blue, and disturbed in red.
%
Error bars were computed  as  Poisson noise.  In this plot, we also present in vertical dashed lines the medians of the color distribution, we can see that for the low-redshift sample, the color density distribution is very similar for the relaxed and disturbed clusters, and the medians are also nearly identical. However, in the bottom panel,  we can see an excess of blue galaxies in the disturbed sample compared to the relaxed clusters. The differences showed in the bottom panel are above the observational uncertainties, and the values of the medians are significantly different, -0.09 for relaxed,   and -0.15 for disturbed clusters. This result  shows that the galaxy population of disturbed clusters  is bluer and possibly less quenched compared to the relaxed clusters.

\section{Discussion and Conclusions}

We used optical data for 84 galaxy clusters detected by SPT-SZ and optically confirmed by the DES survey. The data set is composed by 65 galaxy clusters from DES survey DR2 and 19 high-z clusters from our dedicated observations. Both data sets were observed using the Blanco telescope and the Dark Energy Camera. 
In order to have an homogeneous data set, we performed a calibration using the Stellar Locus Regression code.  This allowed us to correct systematic differences between both data sets.
We divided the final sample in two redshift bins at $z = 0.55$. 
%
The results summarized in Fig.~\ref{Fig:Density}, show that, as expected,  high redshift clusters have a  more significant blue galaxy population with respect to the low redshift subsample  \citep[see e.g.][]{butcher84}. At $z \lesssim 0.55$ the relaxed and disturbed show almost identical CMDs.  However, at $z \gtrsim 0.55$ we see significant differences in their cluster galaxy colour distribution.  Specifically,  the disturbed sample shows an excess of blue galaxies with respect to the relaxed sample.

  This result  provides more evidence that  the galaxy colour distribution not only depends strongly on the global environment \citep[e.g.][]{Peng2010,Iovino2010, Muzzin2012}, but also depends on the dynamical state of the clusters \citep[e.g.,][]{Stroe2017,Zenteno2020a,stroe21}, at least at $z\gtrsim0.55$. 
Our results are consistent with the anti-correlation between the relaxed state of the cluster and their star formation  activity found by \citet{Cohen2015} and \citet{Hou2012}.  The  two possible explanations for the excess in the blue population are the same than the ones proposed by \cite{Cohen2015} for the enhancement of the SFR: {\it i)} the merging process triggers star formation in the galaxies inside the merging clusters generating a more mixed and bluer galaxy population or {\it ii)}  disturbed clusters correspond to a less evolved state than relaxed clusters.  Its worth noting that our results are  consistent with \citet{Pallero2019}.  Using a suite of fully cosmological hydrodynamical simulations,  they showed that most quenched galaxies in massive halos ($M_{200}>10^{14}M_{\odot}$) at $z<0.5$  were quenched in  ex-situ groups or clusters. However for $z>0.5$ the in-situ quenched galaxies dominates their population, suggesting that the galaxies quenched inside the first cluster they fall in. In this scenario, at $z>0.5$ relaxed clusters would have a redder population than disturbed clusters, as they have had more time to evolve their population in-situ.  At $z<0.5$ that difference disappears;  the disturbed sample and relaxed clusters have the same galaxy population since they are being assembled from structures whose members have been already preprocessed within massive substructures. 
%
%
Indeed, this scenario could explain why the galaxy colour distribution at low redshift is the same in both, relaxed and disturbed sample. 
It is worth to mention that our results do not change if SLR correction is not applied to the data in \S\ref{sec:catalogs}.

 Based on the results obtained in this paper, it is not possible to conclude what is the physical mechanism driving the excess of blue galaxies in disturbed high-z clusters. In a follow-up article, we will compare our results against cosmological simulations to characterise the mechanisms responsible for such differences. 
\section*{Acknowledgements}

FA was supported by the doctoral thesis scholarship of Agencia Nacional de Investigaci\'on y Desarrollo (ANID)-Chile, grant 21211648. FAG acknowledges financial support from FONDECYT Regular 1211370. FAG and FA acknowledge funding from the Max Planck Society through a Partner Group grant. FA and FAG gratefully
acknowledges support by the ANID BASAL project FB210003.
ERC is supported by the international Gemini Observatory, a program of NSF’s NOIRLab, which is managed by the Association of Universities for Research in Astronomy (AURA) under a cooperative agreement with the National Science Foundation, on behalf of the Gemini partnership of Argentina, Brazil, Canada, Chile, the Republic of Korea, and the United States of America.
CVM acknowledges support from ANID/FONDECYT through grant 3200918. DHL acknowledges financial support from the MPG Faculty Fellowship program, the new ORIGINS cluster funded by the Deutsche Forschungsgemeinschaft (DFG, German Research Foundation) under Germany's Excellence Strategy - EXC-2094 - 390783311, and the Ludwig-Maximilians-Universit\"at Munich.
\section*{Data Availability}

 The Dark Energy Survey data underlying this article are available at \url{https://www.darkenergysurvey.org/the-des-project/data-access/}



\bibliographystyle{mnras}
\bibliography{library} 




\appendix
\section{Bands Mixture Analysis}
\label{Appendix1}
In this section, we analyze the effect of mixing photometric bands during the construction of the cluster RS. As showed in Table \ref{table:number_clustersri}, we join in the low redshift bin clusters observed with $g-r$ $(z\leq0.33)$ filters and $r-i$ $(0.33<z\leq0.55)$ filters and for the high redshift bin, we join $r-i$ $(0.55<z\leq0.74)$ and $i-z$ $(0.74<z\leq0.9)$ observations. Those bands capture the 4000 {\AA} Balmer break that allows us to detect a prominent cluster Red Sequence. The RS position in the CMD was corrected using the \citet{Bruzual2003} stellar evolution model before stacking to avoid introducing a redshift bias in our sample. However, in this section we repeated the analysis using just the clusters observed between $0.33<z\leq0.74$ using $r-i $ bands to confirm that the differences observed between relaxed and disturbed clusters for high redshift bin is not a consequence of  mixing photometric bands. In Table \ref{table:number_clustersri}, we have the number of stacked clusters for each considered subgroups, we can see that in each subgroup we have at least 12 clusters for low redshift bins and 22 clusters for high redshift bins. 
\begin{table}
\centering
\caption{Number of cluster in low and redshift bins considering the clusters observed using just the $r$ and $i$ bands. }
\label{table:number_clustersri}
\begin{tabular}{ccc}
\hline
Redshift range & Relaxed clusters & Disturbed clusters \\
\hline
$0.33<z<0.55$ & 12 & 22 \\
$0.55\leq z<0.74$ &  19 & 15\\
\hline
\end{tabular}
\end{table}
Similar to the previous analysis, we stacked the CMD in each subgroup after the redshift correction using the \citet{Bruzual2003} stellar evolution model. The morphology of the RS is the same that the showed in the Fig. \ref{Fig:Red_Sequence} and the conclusions are the same: at low redshift bins, the galaxy population is mainly dominated by early-type galaxies, however at high redshift bins, there is a remnant blue (no yet-quenched) galaxy population.

\begin{figure}

\centering
\includegraphics[width=\linewidth, trim={2cm 1cm 2cm 1cm}, clip]{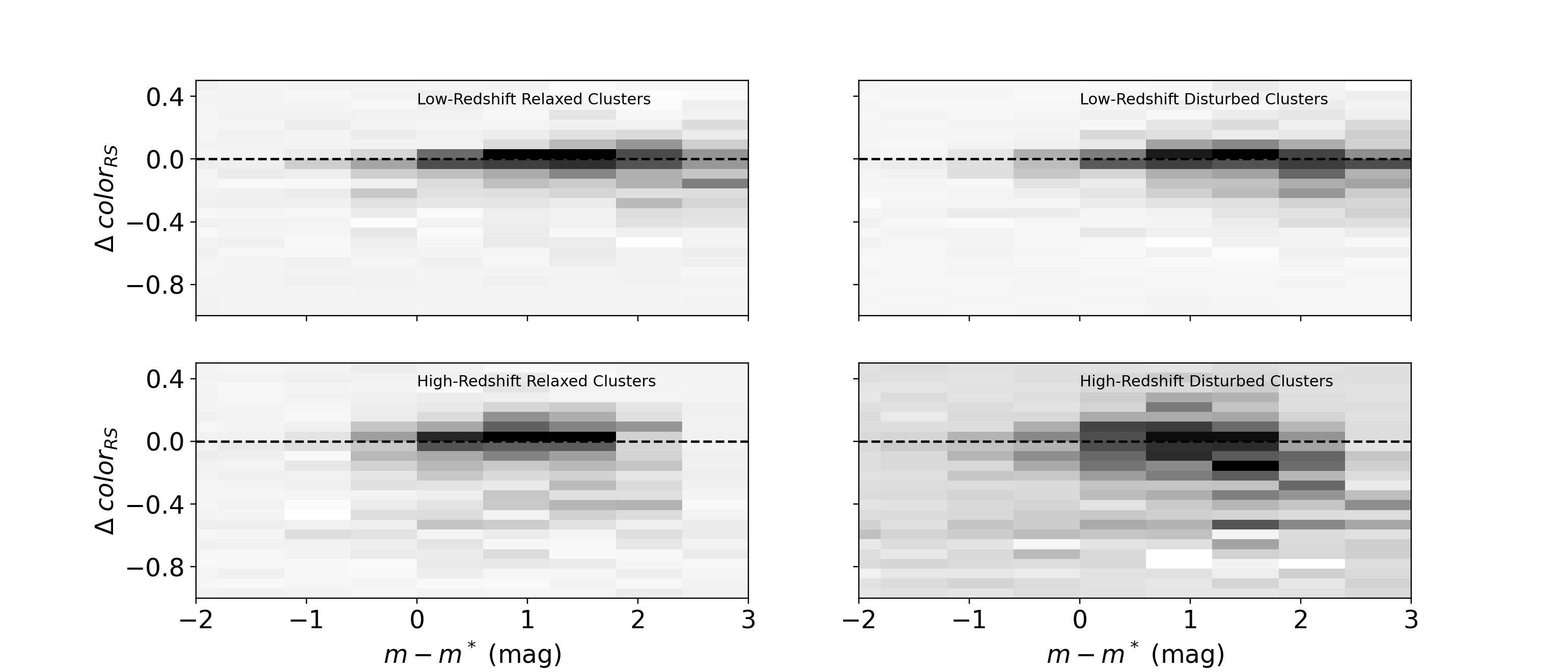}\caption{Similar to Fig.~\ref{Fig:Red_Sequence} but using only the r-i vs i colour magnitude diagram. In the top row we have the low$-z$ $(0.33<z<0.55)$ relaxed (left) and disturbed clusters (right), and in the bottom row we have the relaxed (left) and disturbed (right) clusters for the high$-z$ bin $(0.55<z<0.74)$. It can be seen that combining multiple colours do not affect our results. }
\label{Fig:Red_Sequence_RI}
\end{figure}
\begin{figure}
\centering
\includegraphics[trim={0.0cm 1cm 1cm 2.5cm}, clip,width=0.8\linewidth]{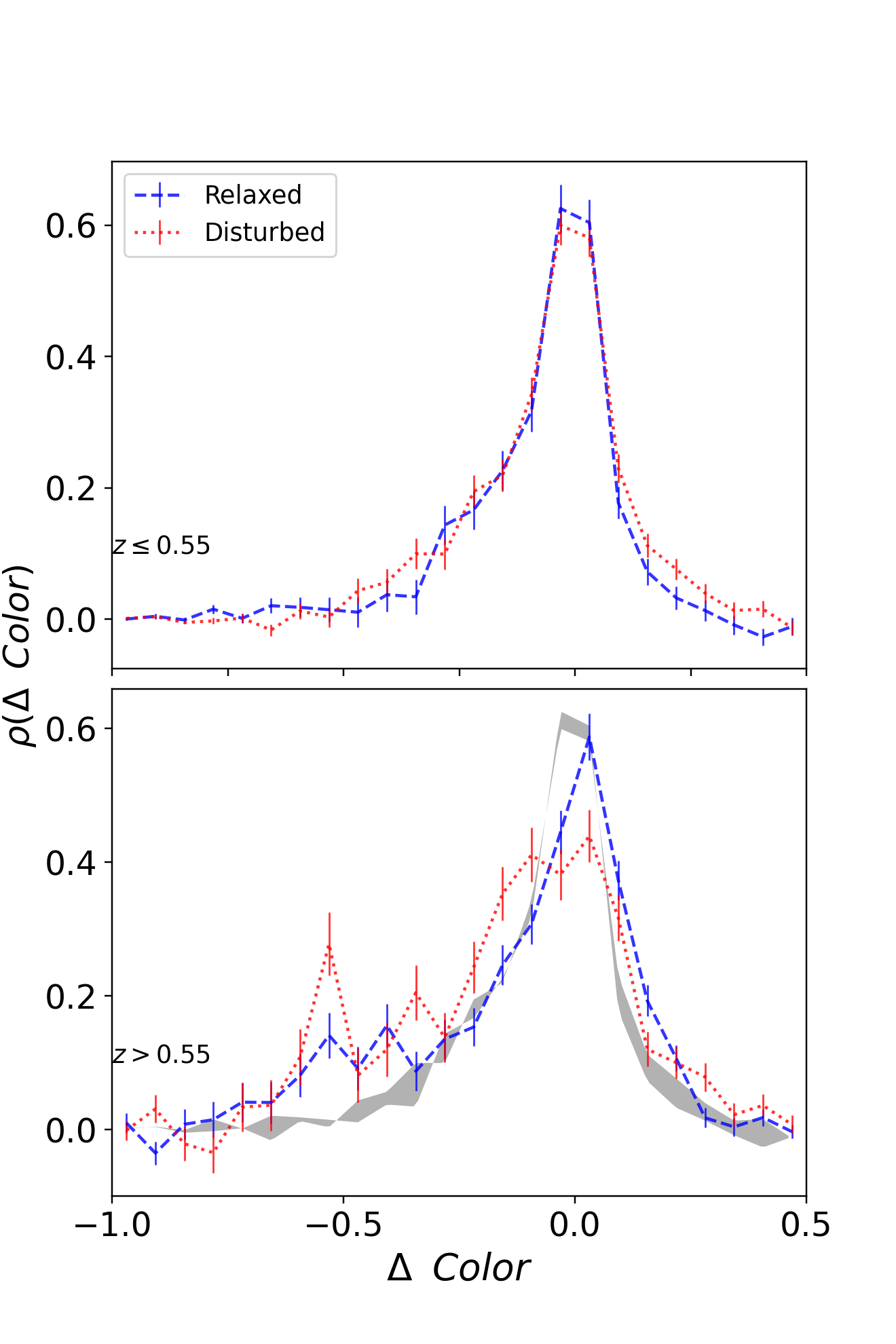}
\caption{Similar to Fig.~\ref{Fig:Density} but using only the r-i vs i colour magnitude diagram. The top panel presents the colour density distribution for the low -redshift bins relaxed in blue line and disturbed in red. Bottom panel the same for $(z>0.55)$. It can be seen that combining multiple colours do not affect our results. }
\label{Fig:Density_RI}
\end{figure}

 
We also repeated a colour density diagram presented in the Fig. \ref{Fig:Density} for cluster observed with $r-i$  bands presenting in the Fig. \ref{Fig:Density_RI}. We can see that those diagrams are very similar and the differences in colour are maintained between relaxed and disturbed clusters at high redshift bins. We also can conclude that the galaxy population at high redshift bins is more mixed in disturbed clusters compared to relaxed ones. With those two plots, we showed that the morphological differences in the RS come from the intrinsic properties of clusters and not from the bands mixing process. 


\section{List of clusters}
In Table \ref{DeROSITA}, we present the list of the 19 high-redshift clusters observed in the frame of the DeROSITAS survey. Alongside their names, we also present the redshift, virial mass (SZ-based), and virial radius obtained using spectroscopic and photometric observations as specified in~\citep{Zenteno2020a}. In Table \ref{DES}, we present the list of the clusters observed by DES including their redshifts, masses and radius. Finally, in Table \ref{DES_obs}, we present the exposure time, FWHM and the number of observations made for one of those 19 high-z galaxy clusters in bands ($r$, $i$, and, $z$).   .  

\begin{table}
    \centering
    \begin{tabular}{c c c c c c}
    \hline
        Name & SPT RA & SPT Dec. & $z$ &  $M_{200}$ & $R_{200}$ \\
SPT-CL & J2000 & J2000 & & $10^{14} h^{-1}_{70}\; M_{\odot}$ & $\arcmin$  \\
\hline
J0014-4952 & 3.69 & -49.87 & 0.752 & 3.27 & 8.10 \\
J0058-6145 & 14.58 & -61.76 & 0.83 & 2.87 & 6.65 \\
J0131-5604 & 22.93 & -56.08 & 0.69 & 3.17 & 6.17 \\
J0230-6028 & 37.64 & -60.46 & 0.68 & 3.06 & 5.41 \\
J0310-4647 & 47.62 & -46.78 & 0.709 & 3.17 & 6.53 \\
J0313-5645 & 48.26 & -56.75 & 0.66 & 2.82 & 3.96 \\
J0324-6236 & 51.05 & -62.60 & 0.75 & 3.21 & 7.57 \\
J0406-4805 & 61.72 & -48.08 & 0.737 & 3.16 & 7.01 \\
J0406-5455 & 61.69 & -54.92 & 0.74 & 2.86 & 5.23 \\
J0422-4608 & 65.74 & -46.14 & 0.7 & 2.79 & 4.36 \\
J0441-4855 & 70.45 & -48.91 & 0.79 & 3.05 & 7.23 \\
J0528-5300 & 82.02 & -53.00 & 0.768 & 2.84 & 5.53 \\
J0533-5005 & 83.40 & -50.09 & 0.881 & 2.64 & 5.78 \\
J2043-5035 & 310.82 & -50.59 & 0.723 & 3.18 & 6.88 \\
J2222-4834 & 335.71 & -48.57 & 0.652 & 3.62 & 8.22 \\
J2228-5828 & 337.21 & -58.46 & 0.71 & 2.85 & 4.77 \\
J2242-4435 & 340.51 & -44.58 & 0.73 & 2.66 & 4.10 \\
J2259-6057 & 344.75 & -60.95 & 0.75 & 3.34 & 8.57 \\  
J2352-4657 & 358.06 & -46.95 & 0.73 & 3.14 & 6.71 \\
\hline
    \end{tabular}
    \caption{19 {\it high-z clusters}  used in this analysis. The redshift, virial mass and virial radius were obtained from \citet{Zenteno2020a} }.
    \label{DeROSITA}
\end{table}

\begin{table}
    \centering
    \begin{tabular}{c c c c c c}
    \hline
Name & SPT RA & SPT Dec. & $z$ & $M_{200}$ & $R_{200}$ \\
SPT-CL & J2000 & J2000 & & $10^{14} h^{-1}_{70}\; M_{\odot}$ & $\arcmin$ \\
\hline
J0000-5748 & 0.24 & -57.80 & 0.702 & 3.25 & 6.91 \\
J0033-6326 & 8.47 & -63.44 & 0.597 & 3.67 & 7.12 \\
J0038-5244 & 9.72 & -52.74 & 0.42 & 4.16 & 4.79 \\
J0107-4855 & 16.88 & -48.91 & 0.6 & 3.08 & 4.24 \\
J0111-5518 & 17.84 & -55.31 & 0.56 & 3.23 & 4.23 \\
J0123-4821 & 20.79 & -48.35 & 0.655 & 3.38 & 6.73 \\
J0135-5904 & 23.97 & -59.08 & 0.49 & 3.57 & 4.28 \\
J0144-4807 & 26.17 & -48.12 & 0.31 & 5.27 & 4.8 \\
J0145-5301 & 26.26 & -53.02 & 0.117 & 14.25 & 7.73 \\
J0147-5622 & 26.96 & -56.37 & 0.64 & 3.01 & 4.54 \\
J0151-5654 & 27.78 & -56.91 & 0.29 & 5.54 & 4.75 \\
J0152-5303 & 28.23 & -53.05 & 0.55 & 3.79 & 6.59 \\
J0200-4852 & 30.14 & -48.87 & 0.498 & 4.18 & 7.13 \\
J0212-4657 & 33.10 & -46.95 & 0.655 & 3.71 & 8.93 \\
J0217-4310 & 34.41 & -43.18 & 0.52 & 3.89 & 6.3 \\
J0231-5403 & 37.77 & -54.05 & 0.59 & 3.26 & 4.87 \\     
J0232-5257 & 38.18 & -52.95 & 0.556 & 4.03 & 8.08 \\
J0243-5930 & 40.86 & -59.51 & 0.635 & 3.48 & 6.92 \\
J0253-6046 & 43.46 & -60.77 & 0.45 & 3.9 & 4.6 \\
J0256-5617 & 44.09 & -56.29 & 0.58 & 3.7 & 6.83 \\
J0257-4817 & 44.44 & -48.29 & 0.46 & 3.93 & 4.95 \\
J0257-5732 & 44.35 & -57.54 & 0.434 & 4.04 & 4.73 \\
J0257-5842 & 44.39 & -58.71 & 0.44 & 4.09 & 5.05 \\
J0304-4748 & 46.15 & -47.81 & 0.51 & 3.93 & 6.2 \\
J0307-5042 & 46.95 & -50.70 & 0.55 & 4.03 & 7.92 \\
J0307-6225 & 46.83 & -62.43 & 0.579 & 3.84 & 7.63 \\
J0317-5935 & 49.32 & -59.58 & 0.469 & 4.12 & 5.96 \\
J0334-4659 & 53.54 & -46.99 & 0.485 & 4.48 & 8.29 \\
J0337-4928 & 54.45 & -49.47 & 0.53 & 3.59 & 5.14 \\
J0337-6300 & 54.46 & -63.01 & 0.48 & 3.77 & 4.81 \\
J0342-5354 & 55.52 & -53.91 & 0.53 & 3.58 & 5.11 \\
J0343-5518 & 55.76 & -55.30 & 0.55 & 3.59 & 5.61 \\
J0352-5647 & 58.23 & -56.76 & 0.649 & 3.34 & 6.41 \\
J0354-5904 & 58.56 & -59.07 & 0.41 & 4.61 & 6.19 \\
J0403-5719 & 60.96 & -57.32 & 0.466 & 4.05 & 5.58 \\
J0429-5233 & 67.43 & -52.56 & 0.53 & 3.32 & 4.08 \\
J0439-4600 & 69.80 & -46.01 & 0.34 & 5.77 & 7.86 \\
J0439-5330 & 69.92 & -53.50 & 0.43 & 4.23 & 5.32 \\
J0451-4952 & 72.96 & -49.87 & 0.39 & 4.34 & 4.6 \\
J0509-5342 & 77.33 & -53.70 & 0.461 & 4.52 & 7.57 \\
J0522-5026 & 80.51 & -50.43 & 0.52 & 3.51 & 4.62 \\
J0526-5018 & 81.50 & -50.31 & 0.58 & 3.15 & 4.25 \\
J0542-4100 & 85.71 & -41.00 & 0.642 & 3.6 & 7.82 \\
J0550-5019 & 87.55 & -50.32 & 0.65 & 2.9 & 4.17 \\
J0551-5709 & 87.90 & -57.15 & 0.423 & 4.79 & 7.42 \\
J0559-5249 & 89.925 & -52.82 & 0.609 & 3.88 & 8.76 \\
J0600-4353 & 90.06 & -43.88 & 0.36 & 5.4 & 7.35 \\
J0611-4724 & 92.92 & -47.41 & 0.49 & 3.9 & 5.56 \\
J0612-4317 & 93.02 & -43.29 & 0.54 & 3.73 & 6.02 \\
J2011-5725 & 302.85 & -57.42 & 0.279 & 5.88 & 5.16 \\
J2022-6323 & 305.52 & -63.39 & 0.383 & 4.85 & 6.17 \\
J2040-5342 & 310.21 & -53.71 & 0.55 & 3.66 & 5.94 \\
J2055-5456 & 313.99 & -54.93 & 0.139 & 11.8 & 6.99 \\
J2130-6458 & 322.72 & -64.97 & 0.316 & 5.96 & 7.25 \\
J2134-4238 & 323.50 & -42.64 & 0.196 & 9.45 & 8.85 \\
J2140-5331 & 325.03 & -53.51 & 0.56 & 3.31 & 4.56 \\
J2146-5736 & 326.69 & -57.61 & 0.602 & 3.36 & 5.57 \\
J2148-6116 & 327.18 & -61.27 & 0.571 & 3.71 & 6.7 \\
J2232-5959 & 338.14 & -59.99 & 0.594 & 3.89 & 8.39 \\
J2233-5339 & 338.32 & -53.65 & 0.44 & 4.81 & 8.23 \\
J2254-5805 & 343.58 & -58.08 & 0.153 & 9.75 & 5.08 \\
J2331-5051 & 352.96 & -50.86 & 0.576 & 3.99 & 8.46 \\
J2332-5358 & 353.10 & -53.96 & 0.402 & 5.08 & 7.89 \\
J2344-4224 & 356.14 & -42.41 & 0.29 & 5.44 & 4.49 \\
J2358-6129 & 359.70 & -61.48 & 0.37 & 4.92 & 5.92 \\
\hline
    \end{tabular}
    \caption{65 DES clusters used in this analysis. The redshift, virial mass and virial radius were obtained from \citet{Zenteno2020a} }.
    \label{DES}
\end{table}

\begin{table}
    \centering
    \begin{tabular}{cccccccccc}
    \hline
Name & \multicolumn{3}{c}{Exposure Times$^a$ (s)} &  \multicolumn{3}{c}{FWHM ('')} & \multicolumn{3}{c}{N$_{obs}$}\\
&$r$&$i$&$z$&$r$&$i$&$z$&$r$&$i$&$z$\\
\hline
J0014-4952 &  --  & 1840 & 2090 &  --  & 0.95 & 0.78 & -- & 10 & 11\\
J0058-6145 &  --  & 3150 & 4746 &  --  & 1.06 & 1.16 & -- & 15 & 21 \\
J0131-5604 &  --  & 1200 & 1800 &  --  & 0.97 & 0.82 & -- & 8  & 10 \\
J0230-6028 &  --  & 1200 & 2938 &  --  & 0.94 & 1.15 & -- & 8  & 13 \\
J0310-4647 &  --  & 3750 & 1800 &  --  & 1.03 & 0.90 & -- & 27 & 10 \\
J0313-5645 &  --  & 1080 & 3080 &  --  & 1.10 & 1.17 & -- & 9  & 14 \\
J0324-6236 &  --  & 2000 & 2640 &  --  & 1.00 & 0.87 & -- & 10 & 15 \\
J0406-4805 &  --  & 1575 & 2460 &  --  & 0.96 & 0.94 & -- & 8  & 12 \\
J0406-5455 &  --  & 1600 & 3750 &  --  & 1.01 & 0.88 & -- & 8  & 15 \\
J0422-4608 &  --  & 2638 & 2152 &  --  & 0.91 & 1.03 & -- & 18 & 13 \\
J0441-4855 &  --  & 1400 & 4200 &  --  & 0.84 & 0.99 & -- & 7  & 14 \\
J0528-5300 &  --  & 3000 & 3300 &  --  & 1.11 & 1.03 & -- & 15 & 11 \\
J0533-5005 &  --  & 5700 & 4815 &  --  & 0.98 & 0.86 & -- & 19 & 15 \\
J2043-5035 & 2045 & 1664 & 1400 & 1.30 & 1.10 & 0.84 & 14 & 8  & 7 \\
J2222-4834 & 1050 & 1080 &  --  & 1.39 & 0.81 &  --  & 9  & 9  & -- \\
J2228-5828 & 600  &  --  &  --  & 1.51 &  --  &  --  & 4  & -- & -- \\
J2242-4435 & 2085 & 1335 & 2000 & 1.22 & 1.52 & 0.81 & 15 & 10 & 10 \\
J2259-6057 &  --  & 3579 & 3600 &  --  & 1.13 & 1.01 & -- & 18 & 18 \\
J2352-4657 &  --  & 1582 & 5350 &  --  & 0.97 & 0.89 & -- & 9  & 22 \\

\hline
    \end{tabular}
    \caption{Observations of the {\it high-z clusters} sample. $^a$The exposure times shown  are in addition to the observations carried out by  DES \citep{abbott21}, which were included in the final  $griz$ stacks.}
    \label{DES_obs}
\end{table}

\label{lastpage}
\end{document}